\documentclass[12pt,floatfix,superscriptaddress,nofootinbib,aps,prc,eqsecnum]{revtex4-2}
\usepackage[dvips]{graphicx}
\usepackage{latexsym,amssymb,amsmath,epsfig,bm,times,psfrag,subfigure}
\usepackage{color}
\usepackage[bookmarksnumbered,bookmarksopen,colorlinks,citecolor=red,linkcolor=blue]{hyperref}
\renewcommand\a{\alpha}
\renewcommand\b{\beta}
\renewcommand\d{\delta}
\renewcommand\k{\kappa}
\renewcommand\l{\lambda}
\renewcommand\r{\rho}

\renewcommand\t{\tau}

\renewcommand\o{\omega}
\newcommand\e{\epsilon}
\newcommand\g{\gamma}

\newcommand\m{\mu}
\newcommand\n{\nu}
\newcommand\x{\xi}
\newcommand\p{\pi}
\newcommand\h{\theta}
\newcommand\s{\sigma}
\newcommand\f{\phi}

\newcommand\ve{\varepsilon}

\renewcommand\O{\Omega}

\newcommand\D{\Delta}

\newcommand\F{\Phi}

\newcommand{\fig}[1]{Fig.~\ref{#1}}
\newcommand{\eq}[1]{Eq.~(\ref{#1})}
\newcommand{\eqs}[2]{Eqs.~(\ref{#1})-(\ref{#2})}

\newcommand\lb{\left(}
\newcommand\rb{\right)}
\newcommand\ls{\left[}
\newcommand\rs{\right]}

\newcommand{\lan}{\langle}
\newcommand{\ran}{\rangle}

\newcommand\pt{\partial}

\newcommand{\bx}{{\bm x}}

\newcommand{\bp}{{\bm p}}

\newcommand{\bv}{{\bm v}}

\newcommand{\bB}{{\bm B}}
\newcommand{\bE}{{\bm E}}

\renewcommand{\part}{{\rm part}}

\renewcommand{\vec}{\boldsymbol}
\newcommand{\be}{\begin{equation}}
\newcommand{\ee}{\end{equation}}
\newcommand{\bear}{\begin{eqnarray}}
\newcommand{\eear}{\end{eqnarray}}
\newcommand{\ba}{\begin{array}}
\newcommand{\ea}{\end{array}}

\usepackage{tikz,xcolor,hyperref}

\definecolor{lime}{HTML}{A6CE39}
\DeclareRobustCommand{\orcidicon}{
	\begin{tikzpicture}
	\draw[lime, fill=lime] (0,0) 
	circle [radius=0.16] 
	node[white] {{\fontfamily{qag}\selectfont \tiny ID}};
	\draw[white, fill=white] (-0.0625,0.095) 
	circle [radius=0.007];
	\end{tikzpicture}
	\hspace{-2mm}
}
\foreach \x in {A, ..., Z}{%
	\expandafter\xdef\csname orcid\x\endcsname{\noexpand\href{https://orcid.org/\csname orcidauthor\x\endcsname}{\noexpand\orcidicon}}
}

\begin{document}

\title{Fluid Acceleration in Heavy-Ion Collisions}
\author{Song-Ze Zhong}
\affiliation{Shanghai Institute of Applied Physics, Chinese Academy of Sciences, Shanghai 201800, China}
\author{Xian-Gai Deng\orcidB{}}
\email{xiangai$_$deng@fudan.edu.cn}
\affiliation{Key Laboratory of Nuclear Physics and Ion-beam Application (MOE), Institute of Modern Physics, Fudan University, Shanghai 200433, China}
\author{Xu-Guang Huang\orcidC{}}
\email{huangxuguang@fudan.edu.cn}
\affiliation{Physics Department and Center for Particle Physics and Field Theory, Fudan University, Shanghai 200438, China}
\affiliation{Key Laboratory of Nuclear Physics and Ion-beam Application (MOE), Institute of Modern Physics, Fudan University, Shanghai 200433, China}
\affiliation{Shanghai Research Center for Theoretical Nuclear Physics, National Natural Science Foundation of China and Fudan University, Shanghai 200438, China}
\author{Yu-Gang Ma\orcidD{}}
\email{mayugang@fudan.edu.cn}
\affiliation{Key Laboratory of Nuclear Physics and Ion-beam Application (MOE), Institute of Modern Physics, Fudan University, Shanghai 200433, China}
\affiliation{Shanghai Research Center for Theoretical Nuclear Physics, National Natural Science Foundation of China and Fudan University, Shanghai 200438, China}
\affiliation{School of Physics, East China Normal University, Shanghai 200241, China}
\date{\today}

\begin{abstract}
We study the generation and space-time evolution of fluid acceleration in heavy-ion collisions using AMPT and UrQMD transport models combined with a Gaussian smearing method. The peak proper acceleration reaches several hundred MeV, with mild model dependence. Transverse acceleration points outward and is strongest at the fireball boundary due to steep pressure gradients and low enthalpy density—a persistent feature even at early times and low energies. Longitudinal acceleration shows strong collision-energy dependence: low-energy collisions exhibit early deceleration from nuclear stopping, while ultra-relativistic collisions produce sharp acceleration pulses from passing nuclei. The volume-averaged acceleration is nearly centrality independent, as extreme acceleration localizes at boundaries. These strong acceleration fields may have important implications for QGP physics, including the Unruh effect mimicking a thermal bath, potential influences on the chiral phase transition and deconfinement, and contributions to spin polarization beyond vorticity.
\end{abstract}
\maketitle

\section {Introduction}\label{intro}
Relativistic heavy-ion collisions provide a unique laboratory for studying quantum chromodynamics (QCD) under extreme conditions \cite{Chen2024,Shou2024}. At sufficiently high collision energies, the energy density deposited in the interaction region is large enough to create a deconfined state of quarks and gluons—the quark–gluon plasma (QGP) \cite{NC2024,Bleicher2024,NC2025,Ma2026}. Experiments at the Relativistic Heavy Ion Collider (RHIC) and the Large Hadron Collider (LHC) have firmly established the existence of this state, whose properties reveal that QCD matter at high temperature behaves as a strongly coupled, nearly perfect fluid \cite{PBM2016,Xu2020,CMS2025}.

In addition to producing QGP, heavy-ion collisions generate extremely strong electromagnetic fields~\cite{Skokov:2009qp,Voronyuk:2011jd,Bzdak:2011yy,Deng:2012pc,Bloczynski:2013mca,Shen2024,Zhao2024,Huang2024}. The magnetic fields in noncentral Au+Au collisions at RHIC energies can reach $ eB \sim m_\pi^2 $, while in Pb+Pb collisions at the LHC it can be an order of magnitude larger~\cite{Skokov:2009qp,Voronyuk:2011jd,Bzdak:2011yy,Deng:2012pc,Deng:2017ljz,Deng:2018frf,Taya:2025utb}. Electric fields of comparable strength arise due to particular nuclear charge distribution or fluctuation~\cite{Bzdak:2011yy,Deng:2012pc,Bloczynski:2012en,Taya:2024wrm,Panda:2024ccj} and from asymmetric collision systems such as Cu+Au~\cite{Hirono:2012rt,Deng:2014uja,Cheng:2019qsn}. When coupled to local $\mathcal{P}$- and $\mathcal{CP}$-odd domains in QGP, these fields induce anomalous transport phenomena, such as the chiral magnetic effect (CME)~\cite{Kharzeev:2007jp,Fukushima:2008xe}, chiral separation effect (CSE)~\cite{Son:2004tq,Metlitski:2005pr}, and chiral electric separation effect (CESE)~\cite{Huang:2013iia,Jiang:2014ura,Ma:2015isa}. Experimental measurements at RHIC and LHC have reported signals qualitatively consistent with some of these theoretical expectations, although the interpretation remains complicated by substantial background contributions; see Refs.~\cite{Liu:2020ymh,Gao:2020vbh,Kharzeev:2020jxw,Li:2025yxx,Shen:2025unr} for recent reviews.

In noncentral heavy-ion collisions, the initial angular momentum of the system can reach $J_0 \sim 10^6-10^7\hbar$, a significant fraction of which is transferred to the produced QGP fireball. This leads to fluid vorticity (local rotation in the fluid) oriented approximately perpendicular to the reaction plane~\cite{Jiang:2016woz,Deng:2016gyh,Wei:2018zfb,Becattini:2015ska,Teryaev:2015gxa,Xie:2016fjj,Ivanov:2017dff,Kolomeitsev:2018svb,Deng:2020ygd,Guo:2021udq,Lei:2021mvp,Xi:2023isk}. Other mechanisms such as inhomogeneous expansion of the fireball~\cite{Becattini:2017gcx,Xia:2018tes,Wei:2018zfb}, jet propagation~\cite{Betz:2007kg,Pang:2016igs}, and strong magnetic field (via the Einstein–de Haas effect) can also create vorticity. Vorticity plays an important role in anomalous transport phenomena such as the chiral vortical effects (CVEs)~\cite{Erdmenger:2008rm,Banerjee:2008th,Son:2009tf}, which generates vector and axial currents along the vorticity direction and may induce baryon number separation and collective excitations such as the chiral vortical wave~\cite{Jiang:2015cva}. It also induce spin polarization of baryons and spin alignment of vector mesons \cite{Chen2023,Chen2025}. Measurements of global hyperon polarization at RHIC and LHC provide compelling evidence for the presence of sizable vorticity in heavy-ion collisions, motivating extensive theoretical and phenomenological investigations; see Refs.~\cite{Huang:2020xyr,Becattini:2020ngo,Huang:2020dtn,Gao:2020lxh,Becattini:2024uha,Niida:2024ntm,Huang:2024ffg} for recent reviews.

From a hydrodynamic perspective, fluid acceleration and vorticity appear on equal footing. The vorticity pseudovector $\omega^\mu$ serves as the natural analogue of the magnetic field $B^\mu$, while the acceleration vector $a^\mu$ plays a role analogous to the electric field $E^\mu$. Although vorticity-induced effects have been extensively studied, fluid acceleration in heavy-ion collisions has received comparatively little attention, despite the fact that the medium undergoes extremely strong acceleration, which drives the rapid hydrodynamic expansion and gives rise to sizable radial flow as well as directed and elliptic flows observed at RHIC and LHC.

Acceleration is of particular interest because of its deep connection to the Unruh effect~\cite{Unruh:1976db}: an observer undergoing constant proper acceleration $a$ perceives the Minkowski vacuum as a thermal state with temperature
\begin{eqnarray}
	\label{unruh}
T_U = \frac{a}{2\pi}.
\end{eqnarray} 
In the context of heavy-ion collisions, it has been argued that partons may experience typical decelerations on the order of the saturation scale~\cite{Kharzeev:2005iz}, $Q_s \sim 1~\mathrm{GeV}$, which corresponds to an effective Unruh temperature comparable to the QCD critical temperature. A recent numerical study using the parton-hadron-string dynamics (PHSD) framework further indicates that, during the early stages of Au+Au collisions at $\sqrt{s} = 4.5$–$11.5~\mathrm{GeV}$, the acceleration of the medium can lead to an Unruh temperature $T_U$ that exceeds the thermodynamic temperature $T$~\cite{Prokhorov:2025vak}. Therefore acceleration is not merely a kinematic feature but a thermodynamic control parameter that could induce novel phase structures in QCD matter. Notably, recent developments have employed Wigner function techniques~\cite{Becattini:2020qol,Palermo:2021hlf} and density operator formulations~\cite{Prokhorov:2018bql,Prokhorov:2019hif,Prokhorov:2019cik,Ambrus:2023smm} to investigate the thermodynamic properties of quantum systems in accelerating frame. In addition, the effects of acceleration on the chiral phase transition and the deconfinement phase transition have become subjects of intense study recently~\cite{Ohsaku:2004rv,Ebert:2006bh,Castorina:2012yg,Benic:2015qha,Casado-Turrion:2019gbg,Basu:2023bcu,Kou:2024dml,Chernodub:2024wis,Chernodub:2025ovo,Zhu:2025pxh,Braguta:2026nfy}. For example, some effective models predict that acceleration restores chiral symmetry (similar to high temperature)~\cite{Ohsaku:2004rv,Ebert:2006bh,Casado-Turrion:2019gbg,Zhu:2025pxh}, while others argue that it acts as a ``refrigerator" enhancing symmetry breaking~\cite{Chernodub:2025ovo,Zhu:2025pxh}. 

The goal of the present paper is to systematically investigate the generation and evolution of acceleration in heavy-ion collisions using the AMPT model~\cite{Lin:2004en} and UrQMD model~\cite{Bass:1998ca,Bleicher:1999xi}. We simulate the spatial and temporal distributions of acceleration in realistic collision geometries, and examine the collision energy dependence of acceleration. These results may provide valuable insights into the thermalization of the QGP, the phase structure of QCD matter along the acceleration axis, and the non-inertial effects in QCD matter.

The paper is structured as follows: In Sec.~\ref{hydro}, we provide a brief review of acceleration in hydrodynamics. In Sec.~\ref{setup}, we describe the setup of the numerical simulations. The numerical results are presented in Sec.~\ref{numer}. Finally, we summarize our findings in Sec.~\ref{summ}. We use natural units where $c = \hbar = k_B = 1$ and the Minkowski metric is $\eta_{\mu\nu} = \rm{diag}(1,-1,-1,-1)$

\section {Review of acceleration in hydrodynamics}\label{hydro}
We consider a relativistic fluid with 4-velocity $u^\m=\g(1,\bv)$ (normalized as $u^2=1$) where $\bv(x)$ is the local 3-velocity and $\g(x)=1/\sqrt{1-\bv^2(x)}$ is the local Lorentz factor. All the information about the fluid motion is contained in the gradient tensor of 4-velocity which can be decomposed as
\begin{eqnarray}
\label{decomu}
\pt_\m u_\n = u_\m a_\n +\sigma_{\m\n}+\o_{\m\n}+\frac{1}{3}\h\D_{\m\n},
\end{eqnarray}
where $\Delta_{\m\n}=\eta_{\m\n}-u_\m u_\n$ is a projector to the directions transverse to $u^\m$, $\h=\pt_\m u^\m$ is the expansion rate of the fluid, $\sigma_{\m\n}=(1/2)(\nabla_\m u_\n+\nabla_\n u_\m-2\h\D_{\m\n}/3)$ is the shear tensor with $\nabla_\m =\D_{\m\n}\pt^\n$, $\o_{\m\n}=(1/2)(\nabla_\m u_\n-\nabla_\n u_\m)$ is the kinematic vorticity tensor, and 
 \begin{eqnarray}
 	\label{def:amu}
 	a^\m = \frac{d u^\m}{d\t} =u^\n \pt_\n u^\m
 \end{eqnarray}
is the 4-acceleration of the fluid with $\t$ the proper time of the fluid element. Therefore, the 4-acceleration is the time-like projection of the 4-velocity gradient representing the curvature of the fluid element's worldline. Note that $a^\m$ is transverse to $u^\m$, $a\cdot u=0$. It is worth writing down the components of $a^\m$ explicitly, $a^\m=\g^2(\g^2\bv\cdot\vec a, \vec a+\g^2\vec a\cdot\bv\bv)$, so that in the non-relativistic limit $a^\m\approx (\bv\cdot\vec a, \vec a)$,  where 
 \begin{eqnarray}
 	\label{def3acce}
 	\vec a=\frac{\pt\vec v}{\pt t}+\vec v\cdot\vec\nabla\bv
 \end{eqnarray}
 is the 3-acceleration of the fluid.
 
The relativistic equation of motion comes from the conservation of the energy-momentum tensor, $ \pt_\mu T^{\mu\nu} = 0 $. Projecting this conservation law orthogonal to the fluid velocity isolates the equation for acceleration (neglecting viscous terms):
\begin{eqnarray}
\label{eoma1}
a^\m = \frac{1}{\ve+P}\nabla^\m P,
\end{eqnarray}
where $\ve$ is the energy density and $P$ is the pressure. This is the relativistic Euler equation showing that for ideal fluid the spatial gradient of pressure accelerates the flow. 

We discuss some interesting consequences related to 4-acceleration $a^\m$. 

(1) It is well known that, in non-relativistic ideal hydrodynamics, when the acceleration is a spatial gradient of some scalar function, vorticity cannot be generated. This is easily seen from the vorticity equation,
\begin{eqnarray}
	\label{nonrelvor}
(\pt_t+{\cal L}_\bv)\bm\o=\bm\nabla\times\bm a,
\end{eqnarray}
where $\bm\o=(1/2)\bm\nabla\times\bv$ is the non-relativistic vorticity 3-vector, ${\cal L}_\bv$ represents the Lie derivative along the stream line. When $\bm a =\bm\nabla \f$ with $\f$ an arbitrary function the right-hand side vanishes and \eq{nonrelvor} thus states that the vorticity is frozen in the fluid flow. This is the local form of the Kelvin's circulation theorem. For relativistic case, if $a^\m$ is a spatial gradient of a scalar function, $a^\m =\nabla^\m \f$ with $\f=\ln w$ ($w$ is a scalar), then the vorticity defined by $\O_{\m\n}= w \o_{\m\n}$ cannot be created. In fact, it is straightforward to derive the following evolution equation for $\O_{\m\n}$,
\begin{eqnarray}
	\label{liebvor}
	({\cal L}_u\O)_{\m\n} =\frac{d \ln w}{d\t}\O_{\m\n}+  w \D^\a_\m\D^\b_\n\nabla_{[\a}a_{\b]},
\end{eqnarray} 
where $A_{[\m\n]}\equiv(1/2)(A_{\m\n}-A_{\n\m})$. When $a^\m =\nabla^\m \ln w$, the right-hand side vanishes. From \eq{eoma1} one finds that this happens when the pressure is solely a function of energy density, $P=P(\ve)$ (borotropic condition) and thus $\f$ is obtained as $\f=\int d P/(\ve+P)$. Therefore, if initially vorticity is zero, it remains zero for ideal barotropic flow. This is the relativistic generalization of Kelvin's circulation theorem.

(2) Though it is natural to expect that it is the vorticity to polarize spin in the QGP fluid, the acceleration can actually also polarize spin. The is a relativistic effect, similar to that for a moving body both the magnetic field $\bm B$ and electric field $\bm E$ can magnitize since the moving body feels an additional induced magnetic field given by $-\bv\times\bm E$ with $\bv$ the moving velocity. In fact, at global equilibrium (or local equilibrium), the so-called thermal vorticity $\varpi_{\m\n}=-(1/2)(\pt_\m \b_\n-\pt_\n\b_\m)$ leads to the following mean spin vector in phase space~\cite{Becattini:2013fla,Fang:2016vpj,Liu:2020flb},
\begin{eqnarray}
	\label{spinpolar}
S^\m(x,\bp)&=&-\frac{s(s+1)}{6 m}\e^{\m\n\r\s}p_\n\varpi_{\r\s} f(1\pm f)\\
&=&\frac{s(s+1)}{6 m T}\e^{\m\n\r\s}p_\n\ls\o_{\r\s}+u_\r\lb a_\s+\frac{\nabla_\s T}{T}\rb\rs f(1\pm f),
\end{eqnarray} 
where $E_p=\sqrt{\bp^2+m^2}$ is the energy of the polarized particle, $p^\m=(p^0=E_p,\bp)$ is the on-shell 4-momentum, and $f(x,\bp)$ is the distribution function with $+$ for bosons and $-$ for fermions. For neutral fluid, the Euler equation (\ref{liebvor}) can be re-written as $a^\m=\nabla^\m T/T$ and thus $\nabla_\s T/T$ in \eq{spinpolar} can be replaced by $a_\s$. To make it more transparent, we write down the mean spin 3-vector in the rest frame of the fluid,
\begin{eqnarray}
	\label{spinpolar2}
	\bm S&=&\frac{s(s+1)}{3 m T}\lb E_p\bm\o+\bp\times\bm a\rb f(1\pm f).
\end{eqnarray} 
This acceleration induced contribution has been proven to be very important for understanding the local spin polarization of hyperons~\cite{Wu:2019eyi,Florkowski:2019voj,Becattini:2021iol,Fu:2021pok}.

(3) Fluid acceleration can also induce a number of transport phenomena. It is well know that acceleration plays the similar role as the temperature gradient to generate the Curie heat current,
\begin{eqnarray}
	\label{heatcur}
	q_C^\m &=&-\k  \lb T a^\m-\nabla^\m T\rb,
\end{eqnarray} 
with $\k\geq 0$ the heat conductivity. In non-relativistic case, it is written as $\bm q_C=-\k(T \bm a+\bm\nabla T)$. The generation of Curie heat current inivertablly creates entropy. Therefore, when the fluid is at local or global equilibrium so that the entropy is maximized, the heat current must vanish. This leads to an important equilibrium condition: $a^\m=\nabla^\m T/T$. For neutral fluid, this is equivalent to the Euler equation (\ref{liebvor}) as it must be; for charged fluid, this further impose constraint for the gradient of chemical potential, $a^\n=\nabla^\n T/T=\nabla^\n \m/\m$ with $\m$ the chemical potential. 

In the presence of magnetic field $\vec B$, the generation of electric current due to the gradient of temperature is called Nernst effect, $\bm J_{N}=\s_N\bB\times\bm\nabla T$. A similar effect can induced by acceleration (the inertial Nernst effect): $\bm J_{N}=\s_N T \bB\times\bm a$ with the same Nernst coefficient as the usual Nernst effect. Under both the gradient of temperature and acceleration, one has the total Nernst current $\bm J_{N}=\s_N\bB\times(\bm\nabla T+T\vec a)$, or in a covariant form~\cite{Hattori:2022hyo}
\begin{eqnarray}
	\label{nernstc}
J^\m_N=\s_N\e^{\m\n\r\s}u_\n B_\r (\nabla_\s T- T a_\s).
\end{eqnarray} 
 Again, at local or global equilibirum, the total Nernst current must vanish which gives that $a^\m =\nabla^\m T/T$.

(4) For a chiral fluid (such as the chiral QGP), acceleration can induce some non-dissipative transport phenomena. For example, the canonical energy-momentum tensor can get a contribution of the form $T^{\m\n}=\l \e^{\m\n\r\s}a_\r u_\s$ with $\l=\m_5(\m_5^2+\p^2 T^2)/(6\p^2)$ for massless Dirac fermions in the so-called thermodynamic frame of velocity and $\m_5$ the chiral chemical potential~\cite{Buzzegoli:2018wpy}.  It is also interesting to notice the following analogue with chiral anomaly. Define the fluid helicity current as $h^\m=\frac{1}{2}\e^{\m\n\r\s}u_\n\o_{\r\s}$ (this is also the kinematic vorticity 4-vector $\o^\m$). Direct calculation gives
\begin{eqnarray}
	\label{fheli}
	\pt_\m h^\m=-2a^\m \o_\m,
\end{eqnarray} 
with $\o^\m=(1/2)\e^{\m\n\r\s}u_\n \o_{\r\s}=h^\m$ the vorticity pseudovector, showing that the acceleration is responsible for the evolution of fluid helicity. Though \eq{fheli} is an identity, it has an interesting anologue with the chiral anomaly equation $\pt_\m j^\m_5=C E_\m B^\m$ with $j_5^\m$ the chiral current and $C=e^2/(2\p^2)$ the anomaly coefficient. If we consider the chiral vortical current $j_5^\m=n_5 u^\m +\xi_5\o^\m$ with $\x_5$ the CVE coefficient~\cite{Erdmenger:2008rm,Banerjee:2008th,Son:2009tf}, we find a generalized chiral anomaly equation that $dn_5/d\t+n_5\h=CE_\m B^\m +2\xi_5 a_\m \o^\m$ (assuming a constant $\x_5$). Its physical meaning becomes more transparent if we re-write it in non-relativistic form, $\pt n_5/\pt t+\vec\nabla\cdot(n_5\bv)=C\bE\cdot\bB+2\xi_5\bm a\cdot\bm\o$. Integrating this equation over space gives $d N_5/dt=C\int d^3\bx\bE\cdot\bB+2\x_5\int d^3\bx\bm a\cdot\bm\o=-(C/2){\cal H}_B+\x_5{\cal H}_v$ with ${\cal H}_B=\int d^3\bx \vec A\cdot\bB$ the magnetic helicity and ${\cal H}_v=\int d^3\bx \bv\cdot\bm\o$ the fluid helicity. This shows that for a chiral fluid the chirality of the constituent particles, the magnetic helicity, and the fluid helicity can be mutually transferred. 

\section {Setup of the numerical simulations}\label{setup}
In this section, we outline the general setup of our numerical simulations. The $z$ axis is chosen along the beam direction of the projectile, while the $x$ axis is aligned with the impact parameter ${\vec b}$ which points from the target to the projectile. The $y$ axis is defined to be perpendicular to the reaction plane (the $x$-$z$ plane). The time origin, $t=0$, is set to the moment when the two colliding nuclei first come into contact in the UrQMD simulation and to the moment of maximum nuclear overlap in the AMPT simulation. However, since the nuclei are highly Lorentz-contracted in AMPT, the difference between the contact time and the time of maximum overlap is negligible.

To compute the 4-acceleration $a^\m$ in transport models like AMPT and UrQMD, we must first numerically define the velocity field. Our approach follows closely Refs.~\cite{Deng:2016gyh,Deng:2020ygd}. We introduce a smearing function $\F(x,x_i)$, where $x$ is the field point and $x_i$ is the coordinate of the $i$-th particle. Using $\F(x,x_i)$, we can smear a physical quantity, such as energy or momentum, carried by the $i$th particle located at $x_i$ to another coordinate point $x$. In this context, $\F(x,x_i)$ can be interpreted as representing the quantum wave packet of the $i$th particle. With $\F(x,x_i)$, we define the particle distribution function $f(x,p)$ as
\begin{eqnarray}
f(x,p)=\frac{1}{{\cal N}}\sum_{i}(2\p)^4\d^{(4)}[p-p_i(t)] \F(x,x_i),
\end{eqnarray}
where ${\cal N}=\int d^3\bx \F(x,x_i)$ is a normalization factor, $p_i^\m=(p_i^0, \bp_i)$ with $p_i^0=\sqrt{\bp_i^2+m_i^2}$, and the summation is over all particles in a given event. The energy-momentum tensor and particle number current are then given by
\begin{eqnarray}
\label{tmunu}
T^{\m\n}(x)&=&\int \frac{d^4 p}{(2\p)^4} \frac{p^\m p^\n}{p^0} f(x,p)=\frac{1}{\cal N}\sum_i\frac{p^\m_i p^\n_i}{p_i^0}\F(x,x_i),\\
\label{jmu}
J^{\m}(x)&=&\int \frac{d^4 p}{(2\p)^4} \frac{p^\m}{p^0} f(x,p)=\frac{1}{\cal N}\sum_i\frac{p^\m_i}{p_i^0}\F(x,x_i).
\end{eqnarray}
Thus, we have two natural ways to define the velocity field for a given colliding event: one is the velocity of energy flow, and the other is the velocity of the particle flow. This is analogous to the Landau-Lifishitz and Eckart velocities, respectively, though in our case $J^\m$ does not necessarily correspond to a conserved current. The velocity of energy current $u^\m=\g(1,\bv)$ ($\g=1/\sqrt{1-\bv^2}$) is the time-like eigenvector of energy-momentum tensor $T^{\m\n}$, $T^{\m\n} u_\n=\ve u^\m$ with $\ve$ the local energy density. This condition leads to
\begin{eqnarray}
\label{energyvelo}
\frac{v^a}{1+(v^a)^2}=\frac{T^{0a}}{T^{aa}+T^{00}},
\end{eqnarray}
with $a=1, 2, 3$ the spatial index. Equation (\ref{energyvelo}) determines $\bv$ and hence $u^\m$. The velocity of particle flow is given by $J^\m=n u^\m$ with $n$ the particle number density. This yields
\begin{eqnarray}
\label{particlevelo}
v^a=\frac{J^{a}}{J^0}.
\end{eqnarray}
Therefore, using \eqs{tmunu}{jmu}, the velocity of energy flow and the velocity of particle flow can be extracted from \eq{energyvelo} and \eq{particlevelo}, respectively. In the following, we will present numerical results mainly based on the velocity of energy flow, while using the velocity of particle flow for a purpose of check.

Different choice for the smearing function $\F(x,x_i)$ lead to different numerical results. In our computations, we adopt a static Gaussian type smearing function,
\begin{eqnarray}
\label{smear2}
\F_{\rm G}(x,x_i)=\frac{1}{\sqrt{2\p\s_z^2}2\p\s_\perp^2}\exp{\ls-\frac{(x-x_i)^2+(y-y_i)^2}{2\s_\perp^2}-\frac{(z-z_i)^2}{2\s_z^2}\rs},
\end{eqnarray}
where $\s_\perp$ and $\s_z$ are two width parameters. This type of smearing function has been widely employed in hydrodynamic simulations~\cite{Pang:2012he,Hirano:2012kj}, as well as in studies of fluid vorticity and hyperon spin polarization. In practical applications of the Gaussian smearing function $\F_G(x,x_i)$ in the construction of the energy-momentum tensor $T^{\mu\nu}$ in \eq{tmunu}, a scaling factor $K$ is introduced to multiply the right-hand side. This factor serves as a phenomenological fitting parameter to reproduce the observed particle multiplicities in hydrodynamic simulations~\cite{Pang:2012he,Hirano:2012kj}. In our AMPT-based calculations, we choose $K = 1.35, 1.45$, and $1.6$ for Au + Au collisions at $\sqrt{s} = 62.4, 200$ GeV, and for Pb + Pb collisions 2.76 TeV, respectively. For calculations based on the UrQMD model, we set $K=1$ for $\sqrt{s}=27$ GeV and below. The remaining parameters are chosen as follows. For both AMPT model and UrQMD model, we take $\sigma_\perp=\sigma_z=0.6$ fm. We have verified that varying these width parameters by up to $50\%$ leads only to minor changes in our numerical results. 


With these settings, we simulate $200$ events for each impact parameter and collision energy below 27 GeV using UrQMD model for Au + Au collisions while $1000$ events for 62.4 GeV and above using AMPT model for Au + Au collisions at RHIC energies (62.4 and 200 GeV) as well as  Pb + Pb collisions at LHC energies (2.76 TeV). The 4-acceleration as well as other physical quantities (e.g., energy density) are then obtained by averaging over all the generated events.

\section {Numerical results}\label{numer}

\subsection {Spatial distribution of acceleration}\label{sec:spa}
\begin{figure}[!t]
	\centering  
		\includegraphics[width=0.9\linewidth]{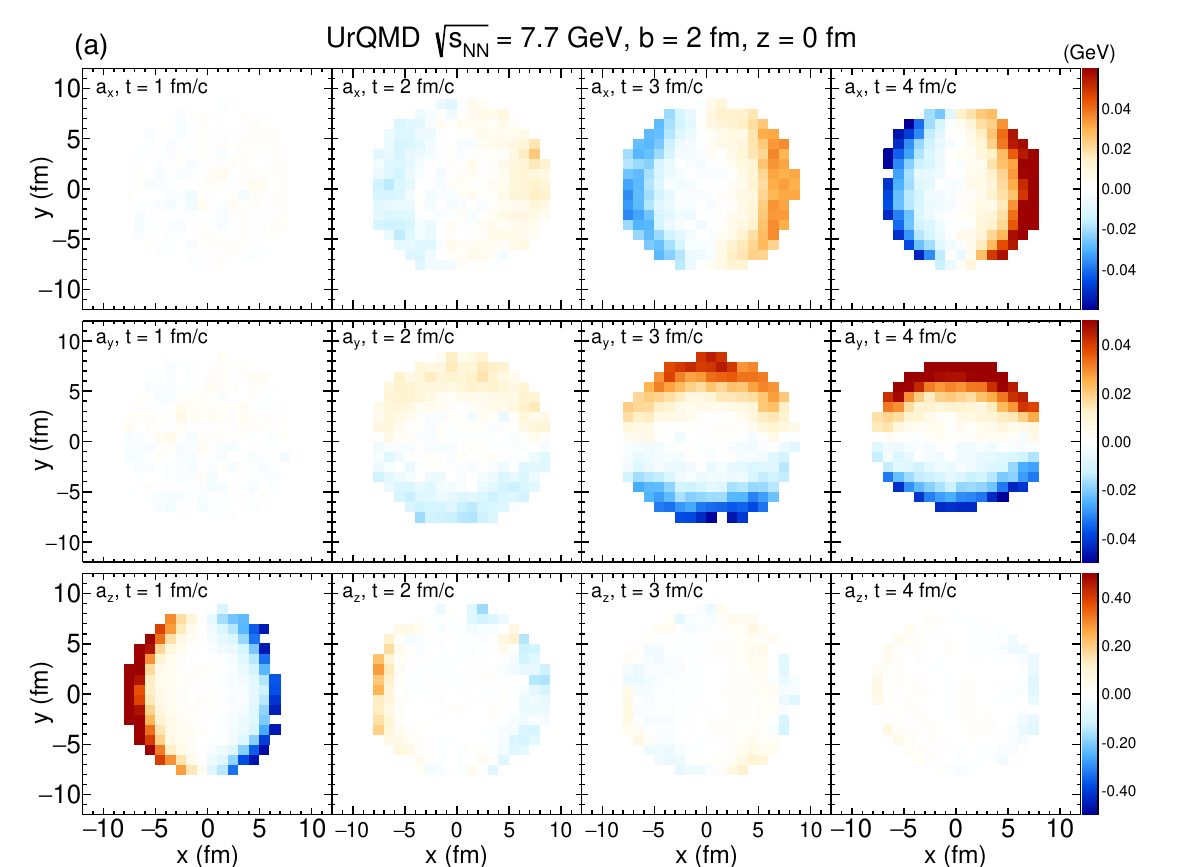}
	\includegraphics[width=0.9\linewidth]{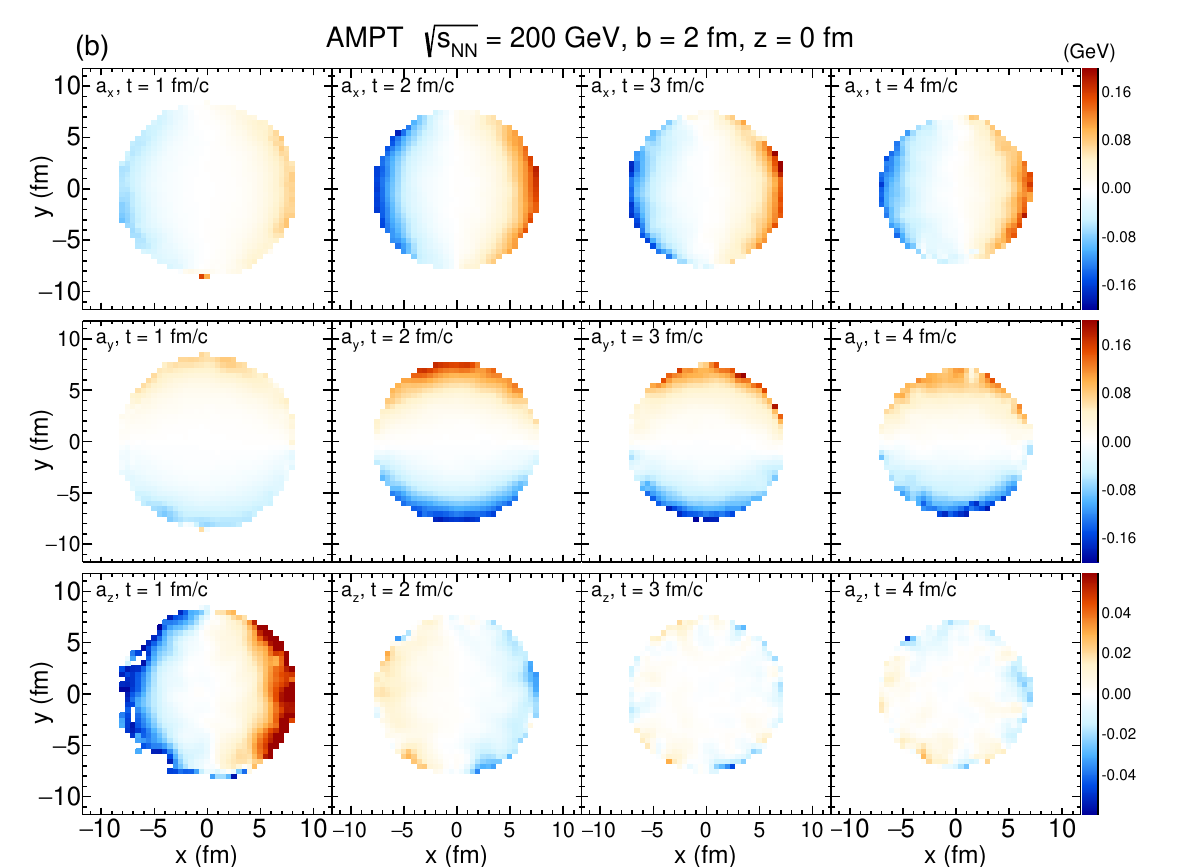}
	\caption{Spatial profiles of the $x, y$, and $z$ components of $a^\m$ in the transverse plane at midrapidity for $\sqrt{s}=7.7$ GeV (upper panel, UrQMD) and $\sqrt{s}=200$ GeV (lower panel, AMPT) for Au + Au collisions.}
	\label{fig:spa}
\end{figure}

We begin by showing the spatial distribution of the 4-acceleration components in the transverse plane at midrapidity ($z=0$) with impact parameter $b=2$ fm. The upper panel of Fig.~\ref{fig:spa} corresponds to $\sqrt{s}=7.7$ GeV simulated using UrQMD model, while the lower panel of \fig{fig:spa} corresponds to $\sqrt{s}=200$ GeV simulated using AMPT model. Note that in these figures $a_x\equiv a^1$, $a_y\equiv a^2$, and $a_z\equiv a^3$ are components of 4-acceleration $a^\m$, rather than the 3-acceleration $\bm a$. In order to isolate the region containing hot and dense medium, following Ref.~\cite{Prokhorov:2025vak}, an energy density threshold of $\ve_c=50$ MeV/fm$^3$ is imposed. Only regions with energy density $\ve>\ve_c$ are included in our analysis. 

As can be seen in \fig{fig:spa}, at both $\sqrt{s}=7.7$ GeV and $200$ GeV the region with $\ve>\ve_c$ shrinks when time grows, reflecting the expansion of the hot medium. 
The transverse acceleration fields, $a_x$ and $a_y$, have a clear outward-pointing behavior that drives the development of radial flow. The magnitudes of the transverse acceleration components $a_x$ and $a_y$ can reach $\sim 150$ MeV for $\sqrt{s}=200$ GeV and $\sim 50$ MeV for $\sqrt{s}=7.7$ GeV at relatively later times (see Sec.~\ref{sec:time} for more details) and robustly concentrated at the peripheral boundary of the fireball. For a thermalized system, this boundary enhancement is understood from the Euler equation, $a^\mu = \nabla^\mu P / (\varepsilon + P)$. Near the boundary of the fireball, the pressure drops abruptly, yielding a steep gradient $\nabla^\m P$, while the enthalpy density $(\varepsilon + P)$ becomes relatively small, further amplifying the acceleration. We note that, unlike Ref.~\cite{Prokhorov:2025vak}, we do not observe an inversion of the acceleration direction in the corona region. 

The longitudinal component, $a_z$, exhibits a highly non-trivial spatial and temporal structure. At $t = 1$ fm/c in the 200 GeV system, $a_z$ is positive in the $x > 0$ hemisphere and negative in the $x < 0$ region. This reflects the forward-backward tilt of the fireball in the reaction plane, leading to a forward (backward) drag via string-tension in AMPT by projectile in the $x>0$ (target in the $x<0$) hemisphere, predominantly localized in the fireball’s peripheral regions; see also the lower-right panel of \fig{fig:tevocomp:mid}. As the system evolves to $t = 2$ and $3$ fm/c, the bulk of the dragged matter has moved away from the midplane ($z=0$ plane), shifting the local pressure maximum to $z>0$ ($z<0$) in the $x>0$ ($x<0$) hemisphere. Therefore, a positive (negative) longitudinal pressure gradient at the midplane in the $x>0$ ($x<0$) hemisphere appears which leads to a negative (positive) $a_z$ at $x>0$ ($x<0$). At later times ($t=4$ fm/c in the figure), the medium moves far away from the $z=0$ plane, so that the acceleration at $z=0$ becomes small and very inhomogeneous. 

At the lower beam energy of $\sqrt{s} = 7.7$ GeV, the overall outward transverse acceleration persists. However, in contrast to the 200 GeV case, the longitudinal acceleration at $z=0$ is always negative (positive) at $x>0$ ($x<0$) and reaches significantly higher magnitudes, exceeding 200 MeV at early times. This behavior reflects the baryon stopping effect, leading to strong initial deceleration near the $z=0$ plane. 

\subsection {Impact parameter dependence}\label{sec:imp}
\begin{figure}[!t]
\centering  
\includegraphics[width=1.0\linewidth]{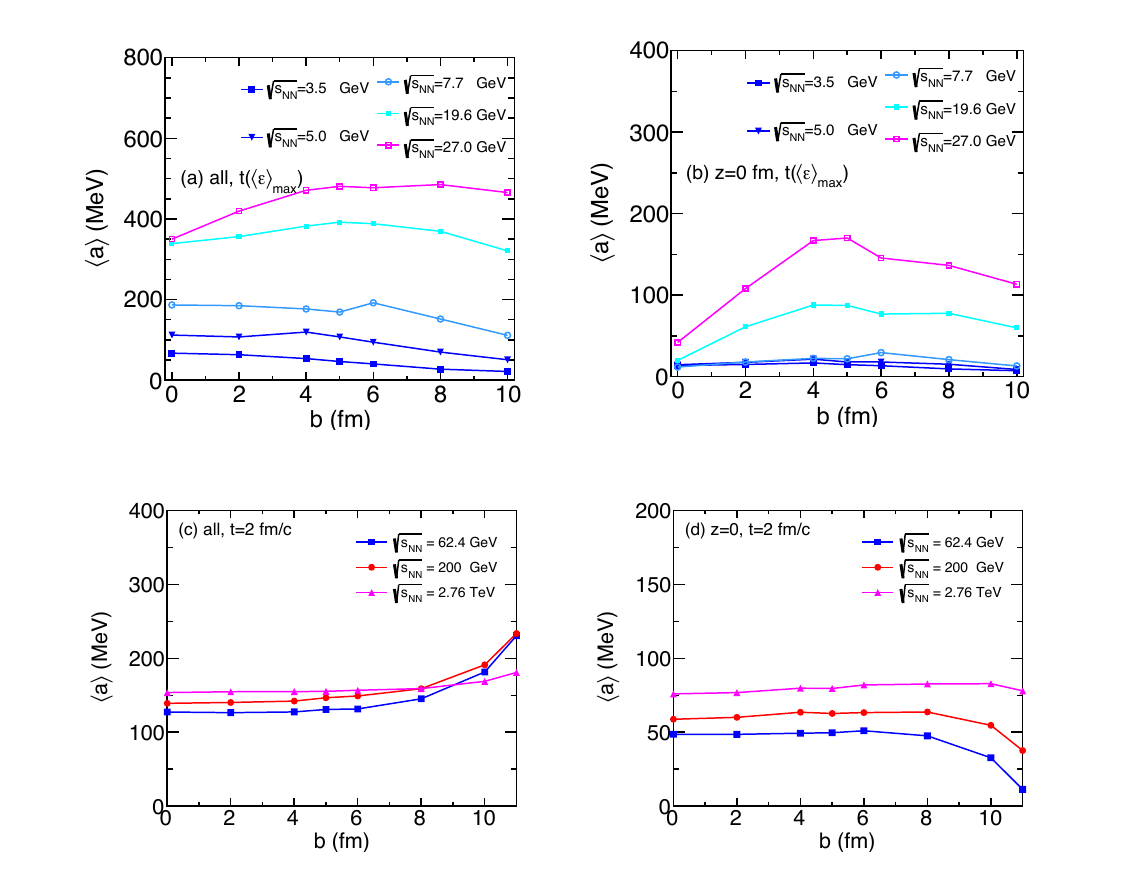}
\caption{Impact parameter dependence of the proper acceleration averaged over the whole volume of the fireball (left panels) and over the transverse plane at midrapidity ($z=0$) (right panels) for lower collision energies (upper panel, UrQMD) and for higher collision energies (lower panel, AMPT). The collision system is Pb + Pb at $\sqrt{s}=2.76$ TeV and Au + Au for other collision energies; the same applies to the figures below.}
\label{fig:impdep}
\end{figure}
In a near-equilibrium system, acceleration is mainly driven by the pressure gradient divided by the enthalpy density. Therefore, both the collision geometry and the thermodynamic state of the created matter influence the local acceleration. In our simulation, though the produced matter may not be thermalized at early times, we still expect the acceleration to exhibit a centrality dependence. To quantify this, we define the proper acceleration as
\begin{equation}
	a = \sqrt{-a^\mu a_\mu}.
\end{equation} 
For a fluid, this Lorentz-invariant quantity measures the magnitude of the 3-acceleration in the local rest frame. 

In \fig{fig:impdep} we show the impact parameter ($b$) (or equivalently, centrality) dependence of the spatially averaged proper acceleration. The upper panels display results for low collision energies simulated by the UrQMD model, while the lower panels show higher energies simulated by the AMPT model. In the left panels, the average is calculated over the whole volume of the fireball,
\begin{eqnarray}
	\lan a\ran=\frac{\int_{\ve>\ve_c} d^3\bx \, a(\bx) }{\int_{\ve>\ve_c} d^3\bx},
\end{eqnarray} 
whereas in the right panels, the average is restricted to the transverse plane at midrapidity ($z=0$),
\begin{eqnarray}
	\lan a\ran=\frac{\int_{\ve>\ve_c} d^2\bx_\perp \, a(\bx_\perp,z=0) }{\int_{\ve>\ve_c} d^2\bx_\perp}.
\end{eqnarray} 
For lower energies in UrQMD simulation, we have chosen the time slice when the averaged energy density reaches its maximum $t(\langle\varepsilon\rangle_{\text{max}})$. For higher energies in AMPT simulation, we have chosen the time slice at a fixed time $t=2$ fm/c. This is because the maximum energy density in the AMPT simulation occurs near the initial time, during which the acceleration fields are dominated by violent, unphysical fluctuations. Because the upper and lower panels are evaluated at different physical stages, their absolute magnitudes cannot be directly compared.

The main features in these spatial averages are the following. First, the global volume averages (left panels) are consistently larger than the local averages at midrapidity (right panels). This indicates that the longitudinal acceleration directed away from the $z=0$ plane remains strong at the selected time slices. 

Second, the proper acceleration increases with collision energy within both the low-energy and high-energy regimes, although the underlying physical mechanisms differ. For lower collision energies, $t(\lan\ve\ran_{\rm max})$ is roughly the moment of maximal compression and the proper acceleration primarily reflects the nuclear stopping power, which naturally grows with the collision energy. In contrast, for higher collision energies ($t=2$ fm/c), the system is closer to local equilibrium. The acceleration is governed by hydrodynamic relation, $a^\mu = \nabla^\mu P / (\varepsilon + P)$. Because both the gradient and the enthalpy density scale up with collision energy, their energies dependencies largely cancel out, resulting in a much milder net increase in acceleration at higher energies. 

Finally, the high-energy results show that the proper acceleration depends very weakly on centrality from central to semi-central collisions. Because the strongest acceleration is localized near the peripheral boundaries of the fireball where the pressure drops sharply to the vacuum, the spatial average over the entire bulk medium remains relatively insensitive to changes in the impact parameter until the collisions become highly peripheral.

\subsection {Time evolution}\label{sec:time}
\begin{figure}[!t]
	\centering  
	\includegraphics[width=1.0\linewidth]{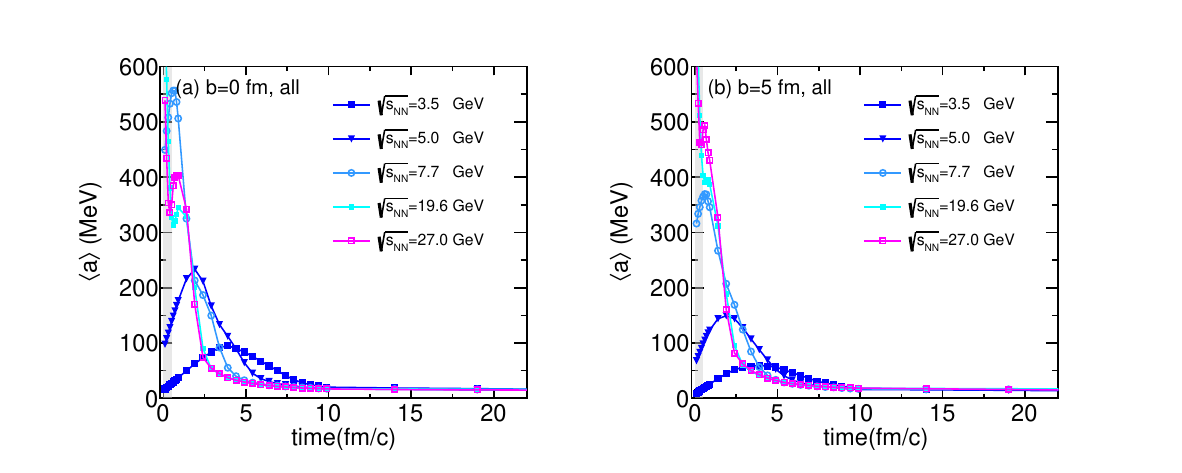}
	\includegraphics[width=1.0\linewidth]{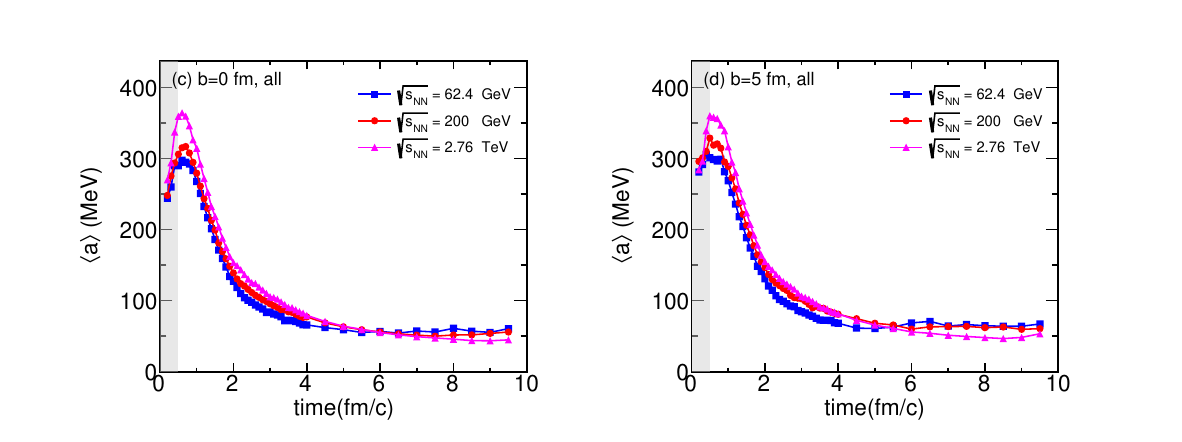}
	\caption{Time evolution of the proper acceleration averaged over the whole fireball for lower collision energies (upper panels, UrQMD) and higher collision energies (lower panels, AMPT) for two centralities, $b=0$ (left panles) and $b=5$ fm (right panels).}
	\label{fig:tevo:who}
\end{figure}
\begin{figure}[!t]
	\centering  
	\includegraphics[width=1.0\linewidth]{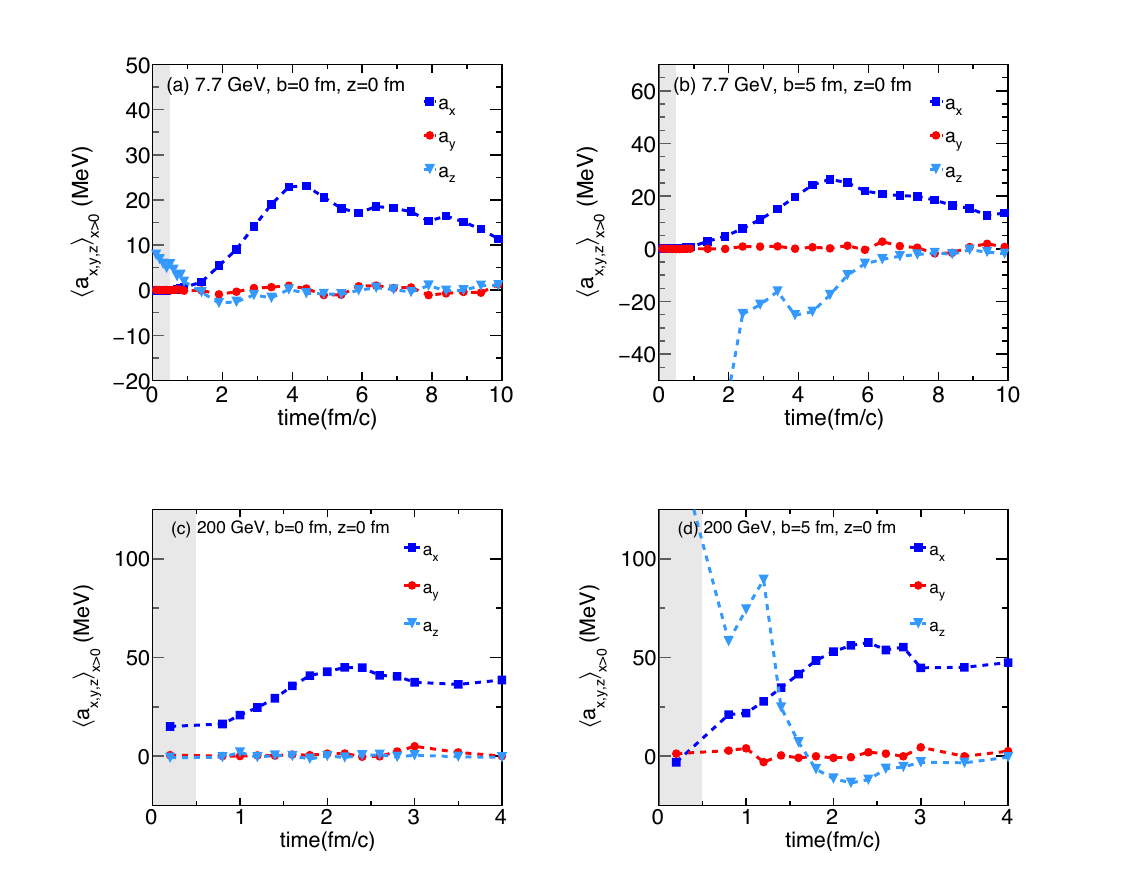}
	\caption{Time evolution of different components of the proper acceleration averaged over the $x>0$ half plane at midrapidity ($z=0$) for $\sqrt{s}=7.7$ GeV (upper panels, UrQMD) and 200 GeV (lower panels, AMPT) for two centralities, $b=0$ (left panels) and $b=5$ fm (right panels).}
	\label{fig:tevocomp:mid}
\end{figure}
In this subsection, we discuss the time evolution of the proper acceleration averaged over the entire fireball volume (defined in the previous subsection). The results, presented in Fig.~\ref{fig:tevo:who}, reveal a highly nontrivial dependence on the collision energy. Focusing first on central collisions ($b=0$) at low collision energies ($\sqrt{s} \sim 3.5$--$7.7$ GeV) simulated by UrQMD, the proper acceleration first increases, reaches a prominent peak, and then decays. The initial rising phase corresponds to the violent compression of the two colliding nuclei, during which the system is strongly decelerated in the longitudinal direction by nuclear stopping. The subsequent decay phase reflects the system's outward expansion. This early-stage deceleration is exceptionally strong, reaching up to $\sim 500$ MeV, but its duration progressively shortens as the collision energy increases. However, we note that at very early times ($t\sim 0$--$0.5$ fm/c, shaded in Fig.~\ref{fig:tevo:who}), the results may strongly depend on the simulation model due to different initial-condition setups. Therefore, results in this period are less physically reliable than those at later times.

Notably, at intermediate energies ($\sqrt{s} = 19.6$ and $27$ GeV), a dip-and-peak structure emerges in the time evolution of the global proper acceleration $\langle a \rangle$. This dip marks the kinematic crossover where the rapidly decaying longitudinal acceleration is overtaken by the building transverse acceleration. At relativistic energies simulated by AMPT (lower panels), the initial longitudinal compression phase is vanishingly short due to extreme Lorentz contraction. Consequently, the global proper acceleration is dominated by a rapid pulse peaking around $t \sim 0.5$--$1.0$ fm/c, which then smoothly decays as the system expands. Interestingly, the peak values of the proper acceleration in the high-energy regime are suppressed compared to the low to  intermediate energy cases. This counterintuitive behavior is perhaps due to that although higher collision energies generate larger absolute pressure gradients ($\nabla^\mu P$), the local enthalpy density ($\varepsilon + P$) scales upward at an even faster rate, leading to a damping of the acceleration comparing to the low energy cases. By integrating over the whole fireball in the high-energy cases, the relatively slow development of transverse radial flow is overwhelmed by the strong longitudinal acceleration at the forward and backward boundaries of the fireball which is insensitive to the collision energies at the relativistic energies, collapsing $\lan a\ran$ into a near-universal time profile. 

To expose the detailed interplay between the longitudinal and transverse dynamics, we show in Fig.~\ref{fig:tevocomp:mid} the spatial components of the 4-acceleration, $\langle a_{x,y,z} \rangle_{x>0}$, averaged over the $x>0$ half-plane at midrapidity ($z=0$). First, due to the up-down symmetry of the collision geometry, $\langle a_y \rangle_{x>0} \approx 0$ for all energies and centralities. Second, in central collisions ($b=0$ fm), exact forward-backward symmetry dictates that the longitudinal acceleration $\langle a_z \rangle_{x>0}$ should identically vanish at $z=0$. However, at $\sqrt{s} = 7.7$ GeV, we observe a small but finite $\langle a_z \rangle_{x>0}$ during the first $\sim 1$ fm/c. This deviation is possibly a numerical artifact; it likely arises from finite grid-size effects and the limited number of simulated events. Third, in non-central collisions ($b=5$ fm), the longitudinal dynamics strongly depend on the collision energy. At $7.7$ GeV, intense baryon stopping generates a strong negative longitudinal acceleration (deceleration) that persists for several fm/c. As the compressed core slowly crosses the midplane, it continues to drive a negative longitudinal acceleration, visible as a broad dip around $t \sim 3$--$5$ fm/c. At $200$ GeV, the Lorentz-contracted nuclei go through each other almost instantaneously. The outgoing projectile remnant drags the formed midrapidity plasma forward via strong string tension at $x>0$. This kinematic wake generates a sharp, positive $\langle a_z \rangle_{x>0}$ pulse around $t \sim 1$ fm/c (consistent with the spatial profiles in Fig.~\ref{fig:spa}). As the bulk of this dragged matter fully crosses the midplane, the local longitudinal pressure gradient inverts, pushing the matter at $z=0$ backward and flipping the acceleration to negative values at later times ($t \gtrsim 2$ fm/c). Finally, across all simulated energies and centralities, the transverse acceleration $\lan a_x \ran_{x>0}$ acts as the universal driver of radial flow, exhibiting a delayed, steady rise to a broad maximum ($t \sim 2-4$ fm/$c$). Eventually, during the expansion of the system, the internal thermal energy is gradually converted into transverse kinetic expansion, driving spatial gradients disperse, and $\lan a_x\ran_{x>0}$ smoothly decays.

\section {Summary and discussions}\label{summ}
In summary, we have systematically investigated the generation and space-time evolution of fluid acceleration in heavy-ion collisions. By utilizing the AMPT and UrQMD transport models, combined with a Gaussian smearing technique to extract the energy-momentum tensor and define the local fluid velocity, we mapped out the local 4-acceleration of the produced medium across a wide  collision energies and impact parameters. 

Our main findings are the following. (1) The peak value of the proper acceleration can be very large, reaching a few hundred MeV for both low and high collision energies. However, we note that its precise value may depend on the model used. (2) In the transverse plane, the acceleration always points outward. The strongest acceleration occurs near the fireball peripheral boundary. From a hydrodynamic point of view (see \eq{eoma1}), this is understood as a two-fold boundary effect: a steep gradient of the pressure and a low enthalpy density. Remarkably, this boundary enhancement persists even at low energies or in very early stages of the evolution. Similar observation was also made in Ref.~\cite{Prokhorov:2025vak}. (3) Longitudinal acceleration shows strong collision-energy dependence. At low energies, nuclear stopping leads to strong early deceleration along the beam direction. At ultra-relativistic energies, Lorentz contraction causes the nuclei to pass through each other quickly. The produced matter at midrapidity is dragged strongly by the passing projectile (target) at forward (backward) directions, producing a sharp acceleration pulse. (4) In low-energy collisions, initial longitudinal stopping dominates early on, but transverse expansion eventually builds up enough strength to overtake it. High-energy collisions bypass this extended deceleration, immediately converting extreme pressure gradients into violent transverse expansion. Interestingly, the averaged proper acceleration depends only weakly on centrality. Because the most extreme acceleration resides at the fireball boundaries, the details of the overlap geometry plays only a minor role in the global volume averaged acceleration.  

These strong acceleration fields may have important implications for quark-gluon plasma physics. Early-stage proper accelerations of hundreds of MeV could mimic a thermal bath through the Unruh effect, suggesting that acceleration might act as a thermodynamic control parameter, potentially influencing the chiral phase transition or the deconfinement properties. Moreover, local fluid acceleration could drive non-dissipative transport phenomena and affect particle spin polarization in ways complementary to fluid vorticity.

\section*{Acknowledgment}
This work is supported by the National Natural Science Foundation of China (Grants No. 12547102 and 12225502), the Natural Science Foundation of Shanghai (Grant No. 23JC1400200 and 23590780100), and the National Key Research and Development Program of China (Grant No. 2022YFA1604900).

\bibliography{Acceleration_in_HIC}
\end{document}